\documentclass[aps,prd,twocolumn]{revtex4}
\tighten
\usepackage{graphics}
\usepackage{graphicx}
\usepackage{epsfig}

\font\BFd=cmmib10 scaled 1200
\font\BFt=cmmib10 scaled 1200
\font\BFs=cmmib10

\def\bb#1{\relax
\ifmmode\mathchoice
{{\hbox{\BFd #1}}}{{\hbox{\BFt #1}}}
{{\hbox{\BFs #1}}}{{\hbox{\BFs #1}}}
\else \mbox{#1} \fi }

\begin{document}

\title{
Gamma-Ray Bursts and Magnetars as
Possible Sources of Ultra High Energy Cosmic Rays: Correlation
of Cosmic Ray Event Positions with IRAS Galaxies}

\author{Shwetabh Singh}
\email[]{shwetabh@astro.berkeley.edu}
\author{Chung-Pei Ma}
\email[]{cpma@astro.berkeley.edu}
\author{Jonathan Arons\footnote{also Physics Department, University
of California,
Berkeley}}
\email[]{arons@astro.berkeley.edu}
\affiliation{Department of Astronomy and Theoretical Astrophysics Center,
University of California, Berkeley,\\ 601 Campbell Hall, Berkeley, CA~94720}

\begin{abstract}

We use the two-dimensional Kolmogorov-Smirnov (KS) test to study the
correlation between the 60 cosmic ray events above $4\times 10^{19}$
eV from the AGASA experiment and the positions of infrared luminous
galaxies from the IRAS PSCz catalog.  These galaxies are expected to
be hosts to gamma ray bursts (GRB) and magnetars, both of which are
associated with core collapse supernovae and have been proposed as
possible acceleration sites for ultra high energy cosmic rays.  We
find consistency between the models and the AGASA events to have been
drawn from the same underlying distribution of positions on the sky
with KS probabilities $\agt 50\%$.  Application of the same test to
the 11 highest AGASA events above $10^{20}$ eV, however, yields a KS
probability of $< 0.5$\%, rejecting the models at 99.5\% significance level.  
Taken at face value, these highest energy
results suggest that the existing cosmic ray events above $10^{20}$ eV
do not owe their origin to long burst GRBs, rapidly rotating
magnetars, or any other events associated with core collapse
supernovae.  The larger data set expected from the AUGER experiment
will test whether this conclusion is real or is a statistical fluke
that we estimate to be at the 2$\sigma$ level.

\end{abstract}

\maketitle

\section{Introduction}

The source of cosmic rays at energies above the
Greisen-Zatsepin-Kuzmin (GZK) feature \cite{GZK}, predicted to appear
at energies above $\sim$ 4 $\times$ 10$^{19}$ eV, remains a mystery
\cite{review}.  These events have been seen by the Akeno Giant Air
Shower Array (AGASA) \cite{AG98}, Fly's Eye \cite{FE93}, Haverah Park
\cite{HP91}, HiRes \cite{HR02}, and Yakutsk \cite{Y91}
experiments. Interaction of charged as well as neutral primaries with
the cosmic background photons through photo-production of pions,
photo-pair production, and inverse Compton scattering \cite{LO94} leads
to a severe energy degradation and constrains the source of a $\sim
10^{20}$ eV charged particle to distances less than $\sim 50$ Mpc (the
high energy GZK zone). The lack of the most commonly hypothesized
accelerators {\it viz.} accretion disks and jets associated with
active galactic nuclei within the GZK zone has led to numerous
alternate proposals for the origin of the Ultra-High Energy Cosmic
Rays (UHECRs).

These explanations fall into two broad categories of top-down and
bottom-up scenarios. Top-down models generally rely on the decays of
very heavy particles, usually remnants from the early universe
\cite{review}.  Bottom-up models are based on the acceleration of
normal charged particles to extremely high energies in astrophysical
objects not otherwise known to possess such effective acceleration
mechanisms \cite{TO02}. These accelerators are expected to show a
correlation with some known class of astrophysical objects in the
local universe.

Supernovae and their remnants have long been suspected as the source
of ordinary cosmic rays below $4\times 10^{19}$ eV.  Two models based
on gamma ray bursts (GRBs) and magnetars for the acceleration of
UHECRs have been proposed that attribute these particles to brief
bursts of particle emission associated with the formation of compact
objects in core collapse supernovae.  In the GRB model, Milgrom and
Usov \cite{mil95}, Vietri \cite{vie95} and Waxman \cite{wax95}
independently suggested that the relativistic shock waves associated
with GRBs might be the acceleration sites for these cosmic rays. The
relativistic shock model has been useful in accounting for the
phenomenology of GRBs \cite{MES02}.  Recent observations of GRB
afterglows have supplied strong evidence that at least the long burst
sources are associated with supernovae occurring in external galaxies
\cite{GRB-SNe}. The prompt association of the afterglow of GRB030329
with SN2003dh gives strong support to the prompt formation of the
collapsed object that drives the outburst, either a black hole, as in
the collapsar model \cite{COL99}, or a rapidly rotating magnetar
\cite{WHE00}; the latter model works only if the jet that drives the
GRB contains only part of the rotational energy delivered, with the
rest driving the supernova envelope, as in the magnetar model for
UHECRs proposed in Ref. \cite{A03}.

Arons \cite{A03} showed that magnetars, a sub-class of neutron stars
with ultra strong surface magnetic fields ($B \sim 10^{15}$ Gauss)
born in core collapse supernovae and hypothesized to have initial spin
rates close to the centrifugal breakup limit, are unipolar inductors
that can accelerate particles in their relativistic winds up to
energies $\sim 10^{22}$ eV.  They have an electrodynamically defined
injection rate sufficient to yield a quantitatively acceptable account
of the observed UHE spectrum, if such events occur in external
galaxies with core collapse supernovae.  Blasi, Epstein, and Olinto
\cite{BLA00} presented a galactic pulsar variant of this idea, and
Milgrom and Usov \cite{mil95} earlier showed that such objects have
voltages (including voltage drops in the wind) large enough to account
for the maximum energies seen in UHECR.  Since some of the magnetars
in our galaxy are physically associated with supernova remnants
\cite{GAEN01}, the association of magnetars with core collapse
supernovae is plausible.

Since both the GRB and magnetar models attribute UHECRs to compact
objects formed in core collapse supernovae (probably in the Ib/c
sub-class), the rate of events that give rise to UHECRs in a galaxy
should be proportional to the supernova rate in galaxies that have
core collapse supernovae.  Likewise, the core collapse supernova rate
has been observed to be strongly correlated to the star formation rate
(SFR) \cite{KE98}, a correlation readily understood from the fact that
such supernovae have short lived, massive stellar progenitors. This
correlation leads us to explore the connection between the UHECR
events and infrared bright spirals that are known to be active star
forming regions.

In this paper we test the hypothesis that the origin of the UHECR
events lies in reasonably prompt acceleration following a core
collapse supernova, by studying the correlations between the angular
locations of UHECR events on the sky and the positions of galaxies
with a high SFR using the two-dimensional Kolmogorov-Smirnov (KS)
test.  We choose the IRAS Point Source Catalogue (PSCz) for our galaxy
sample \cite{IR00}. The catalogue consists of 14800 galaxies with
measured redshifts and far infrared luminosities above a flux limit of
0.6 Jansky at $60 \mu m$.  The sky coverage is $\sim 84 \%$.

Sec~\ref{theory} provides a summary of the acceleration theories.  Sec
\ref{sample} discusses observational and theoretical evidence for the
expected formation rate of the relevant compact objects in infrared
bright spirals, and discusses the use the IRAS PSCz catalogue and the
AGASA sample in our study.  Sec~IV describes the statistical methods
and results on the data and theoretical models. Sec~V and VI provide
further discussion and conclusion of our results.

\section{Compact Object Particle Acceleration Theory \label{theory} }

\subsection{Gamma Ray Bursts}

Gamma ray bursts, and especially their afterglow emission, clearly
reflect the excitation of the magnetized gaseous medium surrounding an
explosive release of energy from a compact object of at least stellar
mass \cite{MES02}.  The observed relativistic expansion rates ({\it
e.g.}, \cite{FRA97}) support that release having given rise to
relativistic outflow, well modeled in the afterglow phase as a
relativistic blast wave propagating through a surrounding
circumstellar or interstellar medium.  In analogy to the apparent
ability of ordinary supernova remnant shocks to accelerate a small
fraction of the particles in the surrounding medium to relativistic
energies, the authors in Refs. \cite{mil95}, \cite{vie95},
\cite{wax95} suppose that relativistic shocks, either that of an
external blast wave expanding with Lorentz factor $\gamma \sim
100-1000$ which creates the afterglow or those of the internal shocks
with $\gamma \sim 2-10$ associated with the prompt GRB, are
responsible for the acceleration of the UHECR.  The luminosity density
for UHECR is $\dot{U}_{\rm UHE} \sim 5 \times 10^{44}$
ergs/Mpc$^3$-sec \cite{A03,VDG03}, about a factor of 5 greater than
the photon luminosity density of GRBs ($\sim 10^{44}$ ergs/Mpc$^3$-sec
\cite{VDG03} but well below the mechanical luminosity density of GRBs,
$\sim 10^{46}$ ergs/Mpc$^3$-sec, assuming $\sim$ 1\% efficiency in
converting internal shock energy into photon energy \cite{SPA00}. The
synchrotron model of GRB emission provides the only evidence for
particle energies in a GRB or its afterglow large enough to correspond
to ions being at relativistic energies.

The energetics of this model are demanding but not exorbitant. The
total observed GRB rate is $n_{\rm g}^{\rm GRB} \nu_{\rm GRB} f_{\rm
b} \sim 0.5 \times 10^{-9} $ Mpc$^{-3}$ yr$^{-1}$ \cite{SCH01}, where
$n_{\rm g}^{\rm GRB}$ is the number density of galaxies hosting GRBs,
$\nu_{\rm GRB}$ is the GRB rate per host galaxy, and $f_{\rm b} =
\Omega_{\rm b} /4\pi$ is the fraction of the outflow beamed into solid
angle $\Omega_{\rm b}$, with $f_{\rm b} \sim 0.01$ inferred in many
afterglows.  Long bursts provide about 70\% of the observed GRB rate
\cite{FI95}, that is, $n_{\rm g}^{\rm SNe} \nu_{\rm GRB} \sim 0.35
\times 10^{-9} $ Mpc$^{-3}$ yr$^{-1}/ f_{\rm b} $, where $n_{\rm
g}^{\rm GRB}$ is the number density of galaxies with core collapse
supernovae $n_{\rm g}^{\rm SNe}$, and $\nu_{\rm GRB}$ is some fraction
of the overall core collapse supernova rate per galaxy.  Then the UHE
energy demanded of each burst source, if long bursts are the sole
contributors, is $\Delta E_{\rm UHE}^{\rm GRB} = \dot{U}_{\rm UHE}
/n_{\rm g}^{\rm SNe} \nu_{\rm GRB} = 10^{54} f_{\rm b} \; {\rm ergs} =
10^{52} (f_{\rm b} /0.01)$ ergs. This energy is somewhat in excess of
the energy put into photons in each burst. As is obvious from the
energy budget of rotation powered pulsars and, to a lesser extent, of
AGN with well developed jets, there is no fundamental difficulty with
particle acceleration power exceeding photon luminosity in a compact
object, especially if the outburst draws its power from the pressure
of large scale electromagnetic fields.

The microphysics arguments advanced for GRBs being a source of UHECR
depends on the theory of particle acceleration in shocks. Most authors
have employed the much studied theory of diffusive Fermi acceleration
in shocks in the test particle limit to estimate maximum energies and
spectra, with the efficiency of shock acceleration left as a free
parameter.  Here, we reiterate the well known remark \cite{KEN84} that
the maximum energy a charged particle can attain in a relativistic
shock wave is $\sim ZeBR_{\rm s} = 3 \times 10^{20} B_{10} R_{\rm
s17}$ eV, where $Z$ is the atomic number of the charged particle, $e$
is the electronic charge, $B_{10}$ is the {\it systematic} magnetic
field of the medium into which the shock propagates in units of 10
Gauss, and $R_{\rm s17}$ is the radius of the shock wave in units of
$10^{17}$ cm; $R_{\rm s17} \sim 1$ is typical of the shock size during
afterglow emission. This energy corresponds to a particle's Larmor
radius having reached the size of the shock, at which point the
particle can move through the motional electric field as a freely
accelerating particle and experience the full electric potential
drop. Acceleration at internal shocks, which occur at much smaller
radii, require concomitantly larger magnetic fields, if GRBs are to be
the acceleration sites for the highest energy particles. Such fields
are {\it much} larger than found anywhere in the interstellar media of
galaxies, forcing the proponents of this model to appeal to
relativistic shock propagation in unusual environments, such as the $B
\propto r^{-1}$ field in the possibly strongly magnetized
circumstellar winds which might surround a presupernova star
\cite{VB88} or the similarly structured field in the relativistic wind
from a newly formed neutron star (Ref. \cite{VDG03} and references
therein) or from a transient magnetized disk around a newly formed
black hole (Ref. \cite{COL99}).

Granted the existence of such environments, the particle spectrum
created by such a shock may be a power law with $f(E) = E^{\rm -p}, \;
p = 2 - 2.5$; detailed studies of the test particle problem yield $p =
2.2-2.3$ \cite{REL}. Models of the acceleration physics which
incorporate some information about the magnetic turbulence that
scatters downstream particles back across the shock front indicate
possibly steeper spectra \cite{OST02}. Such turbulence is an essential
ingredient if the mechanism as applied to shocks with flow Lorentz
factors less than $10^3$ are to accelerate particles to UHECR
energies.  The efficiency with which such shocks can convert flow
energy into accelerated particle spectra is unknown, although judging
from the acceleration efficiencies for electrons and positrons found
in the much more highly relativistic shocks terminating the winds from
rotation powered pulsars, efficiencies as high as 1-10\% appear
possible \cite{A02}.

\subsection{Magnetars}

Magnetars are a special class of neutron stars which form in core
collapse supernovae.  Studies of their association with supernova
remnants in our own galaxy suggest the overall birth rate of magentars
is on the order of 1-10\% of the total core collapse supernova rate
\cite{GAEN01}; this rate is comparable to the rate of type Ib/c
supernovae. Therefore, the magnetar birth rate is in effect
proportional to the supernova rate in a galaxy.  The theory described
in \cite{A03} suggests injection of charged particles with maximum
energy
\begin{equation}
E_{\rm max} = Ze\Phi_{\rm i} = Ze \frac{B_* \Omega_{\rm i}^2}{R_*^3 c^2} =
        3 \times 10^{22} Z B_{15} \Omega_4^2 \; {\rm eV}.
\end{equation}
Here $B_*$ is a magnetar's surface magnetic field, $B_{15} =
B_*/10^{15}$ Gauss, $\Omega_{\rm i} = 10^4 \Omega_4$ s$^{-1}$ is the
initial angular velocity of the neutron star, $R_{*}$ is the stellar
radius, and $c$ is the speed of light. The initial rotation period is
$P_i = 0.64/\Omega_4$ msec.  The ions actually gain their energy in
the relativistic wind electromagnetically expelled from the neutron
star at distances,$r$, much larger than the radii of the star and its
magnetosphere, which allows them to avoid catastrophic radiation
losses; the electric potential in the wind is $rE = rB = \Phi$.  If $
Z = 1-2$, as may be suggested by the shower data \cite{review}, one requires 
$P_{\rm
i} < 2-3$ msec for the model to be viable.  Studies of the termination
shocks of pulsar winds show that the relativistic ions gain their
energy in the wind interior to the winds' termination shocks, rather
than at that (reverse) shock, or at the blast wave driven into the
surrounding interstellar medium \cite{SA03}.  The magnetar model in
\cite{A03} adopts this acceleration site by analogy to pulsars, rather
than appealing to the relativistic blast wave driven by a newly born
magnetar's fields into the surrounding interstellar medium.

The star spins down rapidly, under the influence of electromagnetic
and gravitational wave torques. As it spins down, the voltage and the
particle energies decline. Summing over the formation and spindown
event, one finds a per event injection spectrum proportional to $f(E)
= E^{-1} [1 + (E/E_{\rm g})]^{-1} $ for $E < E_{\rm max} $. Here
$E_{\rm g}$ measures the importance of gravitational wave losses
(calculated for a star with static non-axisymmetric quadrupole
asymmetry) in spinning the star down.  These losses have significance
only at the largest rotation rates, if they matter at all.  When they
exert torques larger than the electromagnetic torque, the star spends
less time at the fastest rotation rates and therefore spends less time
accelerating the highest energy particles, thus causing a steepening
in the spectral slope at the highest energies.  If the star has an
internal magnetic field even stronger than the already large surface
field, equatorial ellipticities $\epsilon_{\rm e}$ in excess of
$10^{-3}$ can exist, in which case $E_{\rm g}$ would be less than
$E_{\rm max}$.  We consider three cases of energy loss due to
gravitational radiation (GR) in the model: no GR loss ($\epsilon_{\rm
e} =0, E_{\rm g} = \infty$); moderate GR loss ($\epsilon_{\rm e} =
0.01, E_{\rm g} = 3 \times 10^{20}$ eV); strong GR loss
($\epsilon_{\rm e} = 0.1, E_{\rm g} = 3 \times 10^{18}$ eV).

The total number of particles injected per event is
\begin{equation}
    N_{\rm i} \approx 2 \frac{c^2 R_*^3 I}{ZeB_*} \approx 
	\frac{10^{43}}{ZB_{15}},
\label{Ni}
\end{equation}
where we have assumed a stellar radius of 10 km and a moment of
inertia $I = 10^{45}$ cgs. In this respect, the magnetar model is much
more specific than the GRB model, since the ion flux is uniquely fixed
by the electrodynamics of the unipolar inductor. In order to tap the
motional EMF of the wind, the particles must cross magnetic field
lines and move parallel to the electric field. In Ref. \cite{A03}, it
was suggested that the current sheets in the structured wind are
unstable to the formation of large amplitude electromagnetic waves,
whose ponderomotive force drives the particles across $B_{\rm wind}$,
thus causing the energy gain, a mechanism analogous to those studied
in the design of advanced accelerators and of some interest for ion
acceleration in the observationally much better constrained pulsar
winds.

The rate at which galaxies inject UHECR into the universe in this
model then is $\dot{n}_{\rm cr} = \nu_{\rm mag}^{\rm fast} N_{\rm i}
n_{\rm galaxy}$, where $n_{\rm galaxy} \approx 0.02$ Mpc$^{-3}$
\cite{bla01}, $N_{\rm i}$ is given by Eq.~(\ref{Ni}), and $\nu_{\rm
mag}^{\rm fast}$ is the birth rate of {\it rapidly rotating} magnetars
per galaxy.  If intergalactic propagation is unaffected by scattering
in an intergalactic magnetic field, multiplication of the source
spectrum $q(E) \propto \dot{n}_{\rm cr} f(E)$ by the energy dependent
GZK loss time yields a spectrum received at the Earth in reasonable
accord with the existing observations of UHECR, as shown in Figure 1
of Ref. \cite{A03}, if $\nu_{\rm mag}^{\rm fast} \approx 10^{-5}$
yr$^{-1}$.  That fast magnetar birth rate lies between 1\% and 10\% of
the total magnetar birth rate inferred for our galaxy, and about 0.1\%
of the total core collapse supernova rate in average star forming
galaxy, $\sim 10^{-2}$ yr$^{-1}$ \cite{CET99}.

In \cite{A03}, the fast magnetar event rate was applied to all normal
galaxies, whose space density is $n_{\rm g} = 0.02$ Mpc$^{-3}$
\cite{bla01}, yielding an event rate $n_{\rm g} \nu_{\rm m}^{\rm fast}
\sim 2 \times 10^{-7}$ Mpc$^{-3}$-yr$^{-1}$.  The luminosity density
of injected energy from the wind, which is about equal to the energy
expended in accelerating the ions, is $10^{46}, \; 0.9 \times 10^{45}$,
and $\; 0.9 \times 10^{43} $ ergs/Mpc$^3 $-yr$^{-1} $, in the no GR,
moderate GR and strong GR cases, assuming the {\it same} $\nu_{\rm
m}^{\rm fast}$ in all three applications of the model.  In fact,
since $\nu_{\rm m}^{\rm fast}$ is an otherwise unknown fraction of the
poorly known $\nu_{\rm m}$, the value of $\nu_{\rm m}^{\rm fast}$ was
adjusted in each case so as to provide the best agreement between the
model's UHECR spectrum and the observations.  The UHECR energy per
event is much smaller than is required for the GRB model, whose
effective rate of occurrence (including beaming) is much less than for
magnetars.  However, there are many normal galaxies within every
resolution element of the arrays used to accumulate the UHECR events,
making angular correlations between all normal galaxies and UHECR
events difficult in testing the magnetar (and the long burst GRB)
hypothesis for the UHECR's origin. The luminous infrared galaxies
offer a better prospect for testing these models, since they are rarer
than ordinary galaxies but still contribute a substantial fraction of
the metagalactic supernova rate (see below).

\section{Study Sample and Sky Coverage\label{sample}}

In this section we discuss how the far infrared luminosity of a galaxy
serves as an indicator of its core collapse supernova rate and why the
IRAS catalogue is chosen as the galaxy sample.  We then discuss the
use of AGASA events as the UHECR sample.

\subsection{Infrared Galaxies}

Type Ib/c and type II supernovae are considered good indicators of the
SFR in a galaxy.  The progenitors of these core collapse supernovae
are young, massive stars that are abundant in active star forming
regions.  Observational support for relating the SFR to the far
infrared luminosity ($\lambda > 20 \mu$m) has come from study of the
IRAS catalogue of $\sim 15000$ infrared bright galaxies.  The method
relies on the fact that a large percentage of the bolometric
luminosity of young stars is absorbed by interstellar dust and then
emitted as thermal radiation from the warmed dust in the far infrared
spectrum \cite{KE98}. The absorption cross section of dust is strongly
peaked in the ultraviolet, where the young stars dominate the
radiation.  This makes the thermal far infrared luminosity of a galaxy
a good proxy for the SFR.

Specifically, we take the supernova rate and therefore the birth rate
of stellar mass collapsed objects to be proportional to the luminosity
of a galaxy at $60 \mu$m.  This wavelength is chosen to reduce the
contamination from the radiation field of older stars at $\alt 10 \mu
m$ as well as the cirrus emission at $ \agt 100 \mu$m due to stellar
optical/near IR radiative excitation of large molecules in the
interstellar medium \cite{KE98}.

Luminous infrared galaxies with high SFRs and supernova rates offer
the best chance of finding a correlation between UHECR event source
directions and sites of core collapse supernovae.  We use the IRAS
PSCz catalogue \cite{IR00} as the host galaxies for GRBs and
magnetars.  This catalogue has a nearly uniform coverage of $\sim
84$\% of the sky, where most of the unmapped area lies within $\sim
10^{\circ}$ north and south of the galactic plane.  The sample
provides redshifts for all galaxies above a flux limit of 0.6 Jansky
at $60 \mu$m.

These galaxies contribute a substantial fraction of the overall core
collapse supernova rate.  From the IRAS catalogue, these objects have
space density $dn_{\rm firg}/dL_{10} \approx 10^{-3} L_{10}^{-1\pm
0.1}$ Mpc$^{-3}$, where $L_{10} = L_{\rm FIR}/10^{10} \; L_\odot $ and
$L_{F\rm IR}$ is the far infrared luminosity \cite{soi86}. The
integral number of these galaxies ($1 < L_{10} < 10^3$) is $n_{\rm
firg} \approx 0.005 $ Mpc$^{-3}$, which should be compared to the
total space density of galaxies, $n_{\rm g} \approx 0.02$ Mpc$^{-3}$
\cite{bla01}; most of these are spirals and other galaxies with some
degree of ongoing star formation.

The general core collapse supernova rate per galaxy, summed over all
galaxy types, is $\dot{S}_{\rm sn} \approx 0.011$ SN per galaxy-year
\cite{CET99}, which yields a volume averaged rate of supernovae of
$\dot{S}_{\rm sn} n_{\rm g} \approx 2.2 \times 10^{-4} $
SN/Mpc$^3$-year. The supernova rate for galaxies with far infrared
emission is $\dot{S}_{\rm sn}^{\rm fir} = 2.5 \times 10^{-4} L_{10}$
SN/(FIR galaxy)-year \cite{MM03}.  Integrating this supernova rate
over the FIR galaxy luminosity function yields $\dot{S}_{\rm sn}^{\rm
fir} n_{\rm firg} \approx 0.7 \times 10^{-4} K_{\rm obsc} (L_{10}^{\rm
max}/300)$ SN/Mpc$^3$-year.  $K_{\rm obsc}$, the correction for
supernovae missed in the existing optical and near infrared supernova
detection surveys, might be as large as 10, and probably is at least
as large as 3 \cite{MM03}. Thus, the luminous infrared galaxies
contribute at least 28\% ($K_{\rm obsc} = 1$) of the total supernova
rate, with a total space density only 25\% that of all normal
galaxies. Furthermore, since the brightest infrared galaxies ($L_{\rm
FIR} > 10^{12} \; L_\odot$) dominate the contribution from all FIR
galaxies, and these are quite rare, with space density $\sim 4 \times
10^{-8}$ Mpc$^{-3}$ \cite{soi86}, correlations of UHECR arrival
directions with the sky positions of the luminous IRAS galaxies offers
a promising opportunity to test the hypothesis that UHECR acceleration
has something to do with core collapse supernovae, as is implied by
the GRB shock and magnetar unipolar inductor models for the
acceleration sites.

\subsection{UHECR Events}

For the UHECR events, we choose the AGASA experiment because among the
ground based arrays, this experiment has the best angular resolution
($1.8^\circ$ FWHM \cite{UCH00}) and provides the largest single
dataset above $4\times 10^{19}$ eV.  The current sample of AGASA UHECR
events contains 60 events above $4 \times 10^{19}$ eV, among which 57
events are published in Ref. \cite{HH00} and 3 recent events with
energies greater than $1 \times 10^{20}$ eV are listed at the AGASA
website \cite{AGWS}.

In order to test whether the AGASA events are consistent with their
origin being from IRAS galaxies, we must first take into account the
different sky coverage by AGASA vs. IRAS and the possible variations
in detector exposure efficiency in AGASA.  Continuously operating
ground-based arrays such as AGASA typically have a uniform coverage in
sidereal time and therefore little exposure variations in right
ascension (RA) \cite{UCH00}.  The detector exposure in declination,
however, depends on the latitude $\theta_0$ of the experiment and
drops to zero beyond certain maximum angle, $\theta_{\rm max}$.  For
AGASA, $\theta_0=35.8^\circ$ and $\theta_{\rm max}=45^\circ$. Beyond $45^\circ$, 
the energy determination of the events becomes uncertain. In terms
of these parameters, the detector efficiency as a function of
declination, $\theta$, is given by \cite{SOM01,ANC03},

\begin{eqnarray}
    \frac{dN_{\rm t}}{d\theta} \propto (\cos\theta_{0} \cos\theta 
	\sin\alpha_{\rm m}
        +\alpha_{\rm m} \sin\theta_{0} sin\theta),
\label{windowt}
\end{eqnarray}

\noindent where $\alpha_{\rm m}$ is given by

\begin{equation}
\alpha_{\rm m} = \left\{ \begin{array}{ll}
   0\,,   & {\rm if}\,\,\,\xi > 1 \\
   \pi\,,  & {\rm if} \,\,\, \xi < -1 \\
   \cos^{-1}\xi \,, \,\, & {\rm otherwise}
\end{array} \right.
\end{equation}

\noindent and

\begin{equation}
\xi\equiv \frac{\cos \theta_{\rm max} - \sin \theta_{0}\,\,\sin \theta}{\cos
\theta_{0}\,\,\,\cos\theta}\,\,.
\end{equation}

For comparison, we find the observed declination distribution published
in Ref.~\cite{UCH00} for the number of AGASA events above $1\times
10^{19}$ eV to be well fit (to $\sim 1\%$) by
\begin{equation}
   \frac{dN_{\rm o}}{d\theta}= \left\{
   \begin{array}{ll}
	-(\theta+10)(\theta-80)/35.5 \,,  \quad & {\rm if}
	 -10^{\circ}<\theta<80^{\circ} \\
	0 \,, &  {\rm otherwise}
    \end{array}   \right.
\label{windowo}
\end{equation}
We have verified that the declination distribution for events above $4
\times 10^{19}$ eV is similar to Eq.~(\ref{windowo}). This is
consistent with the triggering efficiency being uniform above $1
\times 10^{19}$ eV and the UHECR events not showing any appreciable
large scale anisotropy \cite{UCH00,TH99,TA03}.  We use the theoretical
window function in Eq.~(\ref{windowt}) for the statistical tests in
Sec.~IV and discuss the implications of using the observed declination
distribution in Sec.~V.

Fig.~1 shows the sky distribution of the AGASA UHECR events and the
area not mapped in the PSCz catalogue.  Six AGASA events lie in these
unmapped regions and will therefore be excluded, leaving 54 AGASA data
points above $4 \times 10^{19}$ eV for our study.

\section{Statistical Tests \label{tests}}

We use the two-dimensional KS test \cite{P83, FF87} to test whether
the AGASA UHECR events above $4 \times 10^{19}$ eV are consistent with
the cosmic-ray intensity distribution predicted by the GRB and
magnetar models in infrared bright spirals discussed in Secs.~II and
III.  In this section we first describe our calculations of the
expected UHECR distribution from the models and then discuss the
statistical test and the results.

\subsection{Model Predictions}

The spectrum of cosmic rays at Earth depends on both the injection
spectrum at the acceleration sites and possible particle interactions
that would degrade the energies of the primaries between the source
and Earth.  As discussed in Sec.~II, we model the injection spectrum
to be $f(E_{\rm i}) = E_{\rm i}^{\rm -p}, \; p = 2 - 2.5$ for the GRB
models and $f(E_{\rm i})= E_{\rm i}^{-1}[1+(E_{\rm i}/E_{\rm
g})]^{-1}$ for the magnetar models.  For the latter we will explore
the three gravitational radiation cases described in Ref. \cite{A03}:
no GR loss ($E_{\rm g} \rightarrow \infty$), moderate GR loss ($E_{\rm
g}=3\times 10^{20}$ eV), and strong GR loss ($E_{\rm g}=3\times
10^{18}$ eV).  To estimate the energy degradation during propagation,
we use the probability $P_{\rm p}(r,E_{\rm i};E)$ that a proton
created at distance $r$ with energy $E_{\rm i}$ would arrive at Earth
with an energy greater than $E$.  It measures the proton energy
degradation due to the GZK loss processes discussed in Sec.~I.  The
same function was used in our earlier paper \cite{SM03} and has been
calculated using Monte Carlo simulations \cite{FK01} for the
parameters considered here.

We combine the injection spectrum $f(E_{\rm i})$ and propagation loss
$P_{\rm p}$, and calculate the cosmic ray flux at Earth [in (eV m$^2$
s sr)$^{-1}$] by
\begin{eqnarray}
   F(E,\Omega) & = & K(\Omega) \int_{0}^{E_{\rm max}}{\rm d}E_{\rm i} 
\int_{0}^{R_{\rm max}} 
   {\rm d}r[1+z(r)]^{3} \nonumber \\
   & & \times f(E_{\rm i}) \left| \frac{\partial P_{\rm p}(r,E_{\rm 
i};E)}{\partial E}
	\right| \,.
\label{eq:uheflux}
\end{eqnarray}
Here $z$ is redshift, $E_{\rm max}$ is the maximum proton injection
energy, and $K$ is a normalization constant which depends on the cosmic
ray injection model and the viewing direction in the sky,
$\Omega$. The KS test (see next subsection) depends only on the
relative spatial distribution of the UHECR intensity in the sky, we
therefore need not specify $K$ in the theories.

Fig.~2 compares the cosmic ray spectrum $F(E) E^3$ for various GRB and
magnetar models.  For illustrative purposes, the figure is constructed
assuming isotropic injection with a uniform comoving density, i.e.,
$K$ is independent of direction on the sky and is proportional to the
average source density and the event rate per unit time at the present
epoch. If UHECRs come from discrete sources such as galaxies, this
procedure assigns the same cosmic ray luminosity to each source, an
assumption we will relax below.  The curves in Fig.~2 have been
normalized so that all model predictions have the same flux at $4
\times 10^{19}$ eV.  The GZK feature at $E \agt 4 \times 10^{19}$ eV
becomes more pronounced for steeper injection spectra.  (We note that
Fig.~2 corrects a plotting error in the low energy spectrum ($ E <
10^{19}$ eV) of the strong GR case in Fig.~1 of Ref. \cite{A03}.)

Fig.~3 illustrates the expected dependence of the average UHECR flux
$\int_{E_{\rm min}}^{E_{\rm max}} F(E) dE$ on the luminosity distance,
$d_{\rm L}=cH^{-1}_{0}\,(1+z)[z-z^2 (1+q_{0})/{2} + {\cal O}(z^3)]$,
to the host objects for various models: GRBs with a $E^{-2.5}$
injection spectrum (long dashed), magnetars with no GR loss (solid and
dotted), and magnetars with moderate GR loss (short dashed).  Two
lower energy limits $E_{\rm min}$ are shown for comparison for the no
GR loss model.  For $E_{\rm min}= 10^{20}$ eV (dotted), the integrated
flux falls off sharply above $d \sim 50$ Mpc due to the GZK loss
processes.  For $E_{\rm min}=4 \times 10^{19}$ eV (solid), the cutoff
is more gradual and occurs at a larger distance ($d \sim 200$ Mpc).

In order to compare the distribution of UHECR events with the
hypothesized discrete sources, we now need to provide an estimate of
each IRAS galaxy's contribution to the UHECR flux.  We do this by
reinterpreting Eq.~ (\ref{eq:uheflux}) to be the flux expected from
each galaxy in the IRAS catalog, with the galaxies chosen to lie
within the maximum GZK zone indicated by Fig.~3. That is, $K$ in
Eq.~(\ref{eq:uheflux}) is a number chosen to be proportional to the 60
$\mu$m luminosity for each galaxy with a catalogued sky position, all
within the distance $R_{\rm max}$, while the integral over $r$ in
Eq.~(\ref{eq:uheflux}) is carried out for each galaxy separately, from
$0$ to the luminosity distance $d_{\rm L}(z) $ of each galaxy, using the
catalogued redshift, $z$, for that galaxy. This is done for each of
the three model injection spectra of the magnetar model, and for two
possible GRB injection spectra shown in Fig.~2.  Finally, we smooth
the galaxy model's predictions of the UHECR flux as a function of sky
position to the angular resolution of the AGASA experiment, using a
top hat angular smoothing profile.

We include all IRAS PSCz galaxies within $R_{\rm max} = 1$ Gpc,
although our results are not sensitive to this upper cutoff as long as
it is greater than the GZK zone of $\sim$ 200 Mpc.  We estimate
$L_{\rm FIR}$ by using the 60 $\mu$m flux and the flux-luminosity
relationship $L=4 \pi d_{\rm L}^2 f_{\rm FIR}$.  We use matter density
$\Omega_{\rm m}=0.35$, cosmological constant $\Omega_{\Lambda}=0.65$,
and $H_{0}=70\, {\rm km\, s^{-1} Mpc^{-1}}$.  The luminosity distance,
$d_{\rm L}$ differs from linear $d=cz/H_{0}$ by $\sim 15 \%$ at 1000
Mpc and by $\sim 3\%$ at 200 Mpc. Our final results depend very weakly
($\sim 1\%$) on the exact cosmological parameters assumed in the
calculations.

Fig.~4a shows a linearly spaced contour map of the UHECR integrated
flux predicted by our infrared galaxy source model computed from IRAS
galaxies for a magnetar model (no GR case).  The SFRs (and hence
magnetar rates) associated with the galaxies are assumed to be linear
in $L_{\rm FIR}$.  For comparison, all 60 AGASA UHECR events above 4
$\times 10^{19}$ eV are superposed.  Fig.~4b shows the same quantity
for a GRB model with a $E^{-2}$ injection spectrum.  No corrections
for the limited sky coverage and non-uniform detector efficiency in
declination (see Sec~III) have yet been applied.

The question at hand is, do the model and observed angular
distributions of UHECR event rate on the sky significantly differ?
That is, can we disprove the null hypothesis, that the AGASA events
are drawn from sources located in IRAS galaxies?

\subsection{Kolmogorov-Smirnov (KS) Test}

Since the AGASA events do not obviously point towards individual
galaxies, a statistical test is needed to quantify if the AGASA events
are inconsistent with the theoretical UHECR distributions shown in
maps such as Fig.~4.

The two-dimensional KS test is a generalization of the much studied
one-dimensional KS test.  The latter has been discussed extensively in
the literature and analytical proofs for its validity can be found in
\cite{KSA}.  It is based on comparing the cumulative probability
distributions of the sample $H(x)$ and the model $G(x)$. If $D_{\rm
1d}$ is the maximum absolute value of $H(x)-G(x)$ for all $x$, then
the probability $P$ that $H(x)$ and $G(x)$ are drawn from the same
underlying distribution is given by

\begin{equation}
\mathop{lim}_{n\rightarrow \infty} P(>\sqrt{n}D_{\rm 1d})=Q(\sqrt{n}D_{\rm 1d})
\label{prob1d}
\end{equation}

\noindent and

\begin{eqnarray}
Q(\lambda)=2\mathop{\sum}_{k=1}^{\infty} (-1)^{k-1}e^{-2k^2\lambda^{2}}\,,
\label{q}
\end{eqnarray}

\noindent which is valid if the number of data points in the sample,
$n$, is $\agt 80$.  Monte Carlo methods must be used to tabulate the
probabilities for smaller $n$.

Note that a small computed value of $P$ indicates that the null
hypothesis is false, that is, the data did not come from a
distribution represented by the model.  Generally, $P < 0.1$ is
considered reasonable proof that the null hypothesis is false. Larger
values of $P$ simply provide evidence that the data and model are
consistent; the null hypothesis has not been proved false.

The generalization of the KS test to two dimensions is not obvious
because there is no unique cumulative probability distribution in more
than one dimension.  At least two different generalizations of the
test to two dimensions have been proposed \cite{FF87,P83}.  Both
methods have been verified with Monte-Carlo simulations although no
analytical proof for their validity exists.  In this paper we use the
probabilities tabulated in Ref. \cite{FF87}.  The test computes the
{\it difference} between the fraction of data points and the fraction
of theoretical points in each of the four quadrants of the plane of
the sky centered at a data point of angular coordinates
($x_{i},y_{i}$), where the quadrants are defined by
$(x>x_{i},y>y_{i}), (x<x_{i},y>y_{i}), (x<x_{i},y<y_{i}),
(x>x_{i},y<y_{i})$.  This procedure is repeated for all data points,
$i=1,n$, resulting in $4\times n$ numbers for the differences between
the data and model.  The maximum absolute difference value is defined
as $D_{2d}$.  Monte-Carlo simulations are used to obtain the
corresponding probability $P(>\sqrt{n} D_{\rm 2d})$ for a given
$D_{\rm 2d}$ and $n$ \cite{FF87}.  It is also shown in
Ref. \cite{FF87} that the results are almost independent of the
underlying distribution function.  For large $n$, the probability has
an analytic form \cite{PTVF92}:
\begin{eqnarray}
      P(>\sqrt{n} D_{\rm 2d})=Q\left(\frac{\sqrt{n}D_{\rm 2d}}
	{1+\sqrt{1-r_{\rm cc}^{2}} (0.25-0.75/\sqrt{n})}\right)
\label{prob}
\end{eqnarray}
where $Q(\lambda)$ is given by Eq.~(\ref{q}) and $r_{\rm cc}$ is the
linear correlation coefficient of the 2-dimensional data points
$(x_i,y_i), i=1,n$:
\begin{eqnarray}
     r_{\rm cc} \equiv
     \frac{\mathop{\sum}_{\rm i}(x_{\rm i}-\bar{x})(y_{\rm i}-\bar{y})}
	{\sqrt{\mathop{\sum}_{\rm i}(x_{\rm i}-\bar{x})^{2}}
	\sqrt{\mathop{\sum}_{\rm i}(y_{\rm i}-\bar{y})^{2}}}. \,,
\label{defr}
\end{eqnarray}
where $\bar{x}$ and $\bar{y}$ are arithmetic means, and $-1\le r_{\rm
cc}\le 1$.  The value $r_{\rm cc}=1$ is for perfectly correlated $x$
and $y$, which as expected, reduces Eq.~(\ref{prob}) to the same form
as the one-dimensional expression in Eq.~(\ref{prob1d}).  The value
$r_{\rm cc}=0$ is for no correlation between $x$ and $y$, as is the
case for the AGASA events.

In this study, we take the data points to be the angular coordinates
$(x,y)=$(RA,dec) for the 54 AGASA UHECR events above $4\times 10^{19}$
eV that lie within the sky coverage of the IRAS sample.  The model
predictions are the UHECR intensity calculated from IRAS galaxies as
discussed in Sec~IVA, and further weighted by the function $dN_{\rm
t}/d\theta$ in Eq.~(\ref{windowt}) to account for the non-uniform
exposure of the AGASA detector in declination $\theta$.  We determine
the KS parameter $D_{\rm 2d}$ by searching through the four quadrants
centered at the coordinates (RA,dec) of each AGASA event.  For each
quadrant, we compute the difference between the fraction of the UHECR
events that lie in that quadrant, and the fraction of the UHECR
intensity predicted by the models from IRAS galaxies in the same
quadrant.  We then assign the maximal difference from all $4\times 54$
quadrants to be $D_{\rm 2d}$.  We determine the probability $P(>
\sqrt{n} D_{\rm 2d})$ from $D_{\rm 2d}$ by interpolating Table~I of
Ref.~\cite{FF87}.  We do not use the analytic expressions in
Eqs.~(\ref{q}) and (\ref{prob}) that are adopted in the algorithm of
\cite{PTVF92} because they are accurate only for $n \agt 20$ and $P
\alt 0.20$.  (For instance, we find a probability of 32.9\% using the
subroutines provided in \cite{PTVF92} vs. 50.6 \% from \cite{FF87} for
the magnetar-no GR model discussed below.)

\begin{table*}
\caption{ KS probabilities for UHECR events above $4\times 10^{19}$ eV
and above $10^{20}$ eV and the cosmic ray intensity from various
models to be drawn from the same sample.}
\begin{ruledtabular}
\begin{tabular}{ccccc}
        &  $E>4\times 10^{19}$ eV  & $E > 4\times 10^{19}$ eV
        &  $E> 10^{20}$ eV         & $E > 10^{20}$ eV  \\

Model  &  $D_{\rm 2d}(n=54)$          &  $P(>\sqrt{n}D_{\rm 2d})$
        &  $D_{\rm 2d}(n=8)$           &  $P(>\sqrt{n}D_{\rm 2d})$  \\
\hline

Magnetar: no GR loss       & 0.1482 & 50.6 & 0.626 & 0.31\\
Magnetar: moderate GR loss & 0.1472 & 51.7 & 0.615 & 0.39\\
Magnetar: strong GR loss   & 0.1478 & 51.0 & 0.609 & 0.45\\
GRB:  $E^{-2}$             & 0.1475 & 51.5 & 0.626 & 0.31\\
GRB:  $E^{-2.5}$           & 0.1474 & 51.6 & 0.627 & 0.30\\
\end{tabular}
\end{ruledtabular}
\end{table*}

\begin{table*}
\caption{ KS probabilities for UHECR events above $4\times 10^{19}$ eV
and the expected cosmic ray intensity from different SFR models to be
drawn from the same sample.}
\begin{ruledtabular}
\begin{tabular}{ccccc}
          &  Magnetar: no GR & Magnetar: no GR
          &  GRB: $E^{-2}$   &  GRB: $E^{-2}$   \\
    Model &  $D_{\rm 2d}(n=54)$  & $P(>\sqrt{n}D_{\rm 2d})$
	 &  $D_{\rm 2d}(n=54)$  & $P(>\sqrt{n}D_{\rm 2d})$ \\
\hline
SFR$\propto$L$^{0}$ & 0.145 & 55.5 & 0.143 & 57.0\\
SFR$\propto$L$^{0.5}$ & 0.144 & 55.7 & 0.143 & 57.0\\
SFR$\propto$L$^{1.0}$ & 0.148 & 50.6 & 0.147 & 51.5\\
SFR$\propto$L$^{1.5}$ & 0.169 & 24.1 & 0.168 & 26.3\\
SFR$\propto$L$^{2.0}$ & 0.267 & 0.56 & 0.271 & 0.48\\
\end{tabular}
\end{ruledtabular}
\end{table*}

Our results from the KS test for various magnetar and GRB injection
spectra for UHECRs are summarized in Table~I.  The SFR of each IRAS
galaxy is assumed to be proportional to its $L_{\rm FIR}$.  The
consistency between models and data above $4\times 10^{19}$ eV is
generally high with KS probabilities in the range of 50.6\% to 51.7\%
(third column in Table~I); that is, the null hypothesis, that the
angular coordinates of the AGASA events come from the set of angular
coordinates of the IRAS galaxies within the GZK zone, is {\it not}
disproved.  If the data set is restricted to the 8 highest energy
events above $10^{20}$ eV (out of 11 total; 3 lie outside the IRAS
coverage), however, the KS probabilities dive to between 0.30\% and
0.45\% (fifth column); that is, the models are rejected at $> 99.5$\%
significance level.

The assumption of SFR $\propto L_{\rm FIR}$ is reasonable but is
subject to some theoretical and observational uncertainties
(\cite{KE98} and \cite{CET99}).  We therefore test if the KS
probabilities are sensitive to the linear relation of SFR to $L_{\rm
FIR}$ by examining some extreme cases where SFR $\propto L_{\rm
FIR}^\alpha$, with $\alpha=0,0.5,1,1.5$, and 2.  Table~II summarizes
the results for a magnetar (no GR loss) and GRB ($E^{-2}$) injection
spectrum.  The test shows only small variations for $\alpha=0, 0.5$,
and 1, with the probabilities dropping sharply only for very steep power laws to
24.1\% for $\alpha=1.5$ and 0.56\% for $\alpha=2$ for the magnetar (no
GR loss) injection spectrum.  Similar results are found for the GRB
injection spectrum.  We conclude from this test that our results
are robust to all reasonable range of $\alpha\sim 1$.

Starburst galaxies are generally associated with luminous far-infrared
galaxies ($L_{\rm FIR} \agt 10^{11}L_{\odot}$) \cite{MM03}.  To explore a
connection between these galaxies and the UHECR events, we separate
all IRAS galaxies within 200 Mpc into two samples with $L_{\rm FIR}$ above
and below $10^{11}L_{\odot}$.  We find the KS probability to be 44.3\%
and 60.1\% for the above and below $10^{11}L_{\odot}$ samples,
respectively.  Starburst galaxies are therefore not ruled out as
possible UHECR sites, although the KS probability does not
favor starbursts more than ordinary IRAS galaxies.  Combining the two
samples gives a probability of 49.6\%, which is very close to 50.6\%
discussed above for the galaxy sample out to 1 Gpc. This is to be
expected as most contribution to the UHECR flux comes from the GZK
zone as indicated in Fig.~3.

We perform the KS test on a number of additional toy models {\it vs.}
AGASA UHECR events to explore the discriminating power of the KS
method on model assumptions.  In all cases we make sure to weight the
theoretical predictions with the same AGASA detector exposure function
$dN_{\rm t}/d\theta$ for a fair comparison.  In the first toy model,
we assume an isotropic distribution of cosmic ray intensity and obtain
a probabilities of 46.2\% and 9.1\% for the UHECR events above $4 \times 
10^{19}$ and $10^{20}$ eV respectively to have been drawn from
an isotropic distribution.  In the second toy model, we ignore the
redshift and $L_{\rm FIR}$ of the IRAS galaxies and simply test their
sky positions vs. the 54 AGASA events.  We find a KS probability of
54.5\% for UHECR events above $4 \times 10^{19}$ eV and 10.9\% for events above 
$10^{20}$ eV.  In the third toy model, we take the sky positions of the 2702
GRBs in the Burst And Transient Source Experiment (BATSE) catalog
({http://www.batse.msfc.nasa.gov/batse}).  Since the sky coverage here
is 100\%, we use all 60 AGASA UHECR events in this test.  We obtain a
KS probabilities of 42.3\%(E$>4 \times 10^{19}$ eV) and 5.2\%(E$>10^{20}$ eV). A 
realistic calculation of the correlation between the BATSE GRBs and UHECR events 
from AGASA gives a negative result. \cite{EFS03} The numbers from these three 
toy models are
consistent with the fact that IRAS galaxies, BATSE GRBs, and the AGASA
events are all nearly isotropically distributed on the sky on large
angular scales \cite{SUT99,BRI96,UCH00}.  We do emphasize that taking
only the sky positions of source catalogs and correlating them with
AGASA coordinates (as done in these toy models and occasionally in the
literature) is naive since it ignores important energy loss processes
of the primaries during their propagation and possible dependence of
the UHECR flux on source properties such as their intrinsic
luminosity.

\section{Discussion}

In this paper we have found using the two-dimensional KS test that the
AGASA data for the UHECR events with $E > 4\times 10^{19}$ eV are
consistent (at $\sim 50\%$ significance level) with the expected cosmic
ray intensity distribution in the sky predicted by models in which the
acceleration mechanism for the primaries resides in GRBs or magnetars
in infrared bright galaxies.  When only the highest UHECR events ( $E
>10^{20}$ eV) are used, however, we find the KS probabilities to be
$<0.5 \%$, indicating the models are rejected at $>99.5 \%$ significance
level.  When the energy cut on the UHECR sample is varied to $E>
5\times 10^{19}$ eV, $> 6.3\times 10^{19}$ eV, and $>8 \times 10^{19}$
eV, we find the corresponding KS probabilities to be 54.2\%, 6.73\%,
and 0.92\% for the magnetar model (no GR loss), suggesting a better
correlation of the UHECR events at lower energies.

To establish whether the low KS probabilities at the highest energies
are due to the paucity of data points or due to the failure of the
models under consideration, we draw 100 random samples of 8 events
each from the data in the energy range, $4 \times 10^{19}$ eV $<$ E
$<10^{20}$ eV.  There are 46 events in this energy range in the region
mapped by IRAS and the KS probability is 43.3\% for the magnetar with
no GR version of the model.  For a fair comparison with the AGASA
data, we consider only those samples for which the absolute value of
the linear correlation coefficient $|r_{\rm cc}|$ defined in
Eq.~(\ref{defr}) is less than 0.1.  We find that the mean KS
probability of the 100 random samples is 52.8 \% with a standard
deviation of 24.4 \%.  It is therefore unlikely (at $\sim 2\sigma$
level) that the poor agreement between the data above $10^{20}$ eV and
the models considered here is due to the small number of AGASA events
above $10^{20}$ eV.  The AUGER experiment \cite{AUGER} with a
projected $\sim 60$ events per year above $10^{20}$ eV \cite{DOV01}
will help resolve whether these highest AGASA events indeed represent
a $2\sigma$ statistical fluke.  Taken at face value, however, the
existing UHECR data suggest that at least the cosmic ray events above
$10^{20}$ eV do not owe their origin to long burst GRBs, rapidly
rotating magnetars, or any other events associated with core collapse
supernovae.

Our work can be contrasted with a previous work \cite{SGM02} on the
correlations between the UHECR events and luminous infrared galaxies,
in which the authors used the two-dimensional generalization of the
Smirnov-Cramer-von Mises test.  The conclusions of the two papers are
in broad agreement, although our significance levels differ from theirs
partially because unlike the K-S test used here, the generalization of
the Smirnov-Cramer-von Mises test to two dimensions used in
Ref.~\cite{SGM02} does not yield a unique significance level and there
is a large scatter.  Moreover, Ref.~\cite{SGM02} does not exclude the
AGASA data points that lie in the unmapped region of the IRAS survey.  To 
illustrate the importance of correcting for
the detector exposure, we replace the window function in
Eq.~(\ref{windowt}) with a tophat function that is equal to 1 for
$-10^{\circ}<\theta<80^{\circ}$ and zero otherwise, and find the
KS probability to drop to 33.7\% from 50.6\% (for the magnetar model
with no GR loss).

We also test the robustness of our results by using the observed
window function given by Eq.~(\ref{windowo}) instead of the
theoretical function in Eq.~(\ref{windowt}).  We find this to have
only a $\sim 10$\% effect, changing the KS probabilities to 55.7\%,
57.5\%, 45.2\%, 42.8\%, and 43.1\% for the five injection spectra
(from top down) listed in Table~I.

The significance of our comparison of the angular coordinates of the
UHE events on the sky to the coordinates of the IRAS galaxies relies
on rectilinear particle paths through intergalactic space.  Could
scattering in intergalactic magnetic fields affect our conclusion?  An
interesting constraint on the intergalactic field and the fluctuation
correlation length from the GRB model is provided by the 8 doublet and
2 triplet UHECR events occurring with $3^{\circ}$ of each other on the
sky \cite{UCH00}, among a total of 92 events with $E > 4\times
10^{19}$ eV that have been observed over the last 40 years.  Since no
such close pairs of GRBs have been observed, each UHECR doublet and
triplet are therefore events from the single GRBs.  However, these
events are separated by a decade or more, yet GRBs last dynamically
only for a few months.  The particles from a GRB therefore must have
undergone time delays $> 10$ years during their propagation through
the intergalactic medium, an effect which requires some form of
scattering along the propagation path.  The same considerations apply
to the magnetar model, although in contrast to GRBs, photon
observations of magnetar formation do not exist, so we lack explicit
observational evidence against multiple magnetars occurring within 10
year time intervals at the same angular position.

In addition, GRBs within the very high energy GZK zone are extremely
rare. The GRB rate (long bursts only) is $\sim 3.5 \times 10^{-10}$
Mpc$^{-3}$ yr$^{-1}$. Particles with energy above $4 \times 10^{19}$
eV must have been emitted within the GZK zone, $D < D_{\rm GZK} = 100
E_{20}^{-2}$ Mpc ($3 > E_{20} > 0.4$), assuming straight line
propagation. Here $E_{20}$ is the energy in units of $10^{20}$ eV.
Then an average of $\sim 700 E_{20}/10^6$ years pass between each
separate GRB contribution to the flux of such extremely high energy
particles observed at Earth. Therefore at the highest energies, the
contributions from the very rare nearby bursts must be spread out in
time, if the observed event rate is to be attributed to this source.
The magnetar model, with its larger event rate, does not require
temporal dispersion of each burst of particle injection in order to
account for the events observed above $10^{20}$ eV.  Such dispersion
is required, however, in order to account for the doublets and
triplets, since each magnetar injection event lasts only a few days
\cite{A03}

The most likely origin for such time delays is charged particle motion
in an intergalactic magnetic field of average strength,$B_{\rm igm}$,
and irregularities with amplitude $\delta B \sim B_{\rm igm}$ and
correlation length $l_{\rm c}$.  One can show, by simple random walk
arguments, that since a particle is deflected by the small angle
$\delta \theta \approx l_{\rm c}/r_{\rm L} = (ZeB_{\rm igm} l_{\rm c}
/E) \ll 1$ in crossing a single region of correlated field, the
particles from an impulsive event at distance D have arrival angles
smeared over an angular width $\Theta = (Dl_{\rm c}/r_{\rm L}^2)^{1/2}
=0.4^\circ l_{\rm Mpc}^{1/2} ZB_{\rm i10} /E_{20}^2 (D/D_{\rm
GZK})^{1/2} $.  Here $B_{\rm i10} = B_{\rm igm} /10^{-10}$ Gauss and
$l_{\rm Mpc} = l_{\rm c}/1$ Mpc.  Simple geometry shows the time delay
associated with this angular scattering to be $\Delta T_{\rm d} =
(D/2c) \Theta^2 = 365 (D/D_{\rm GZK})^2 (Z^2 B_{\rm i10}^2 /E_{20}^6)
l_{\rm Mpc}$ years.  At the same time, the scattering cannot smear the
inferred angular position of the source by an amount substantially
larger than the angular resolution of the AGASA array.  The
requirements $\Theta < 3^\circ$ and $\Delta T_{\rm d} > 10$ years
constrain the intergalactic field and its fluctuation correlation
length to lie in the range $8 E_{20}^2 (D_{\rm GZK}/D)^{1/2} > ZB_{\rm
i10} l_{\rm Mpc}^{1/2} > 0.04 E_{20}^3 (D_{\rm GZK}/D)$, $0.4 < E_{20}
< 3$, if an impulsive event model is to be a viable candidate for the
origin of UHECR.  See Ref. \cite{BD03} for a complete analysis of this
issue, including a discussion of likely values for the intergalactic
field and the effect of transport through galaxy clusters as well as
the general intergalactic medium.

To estimate how much magnetic fields can effect our results, we take
all 60 AGASA data points above $4 \times 10^{19}$ eV and randomly
displace them by $3^{\circ}$.  After rejecting the 6 events that fall
in the unmapped region of IRAS, we find a KS probability of 46.3 \%
for the no GR magnetar model, a result consistent with our earlier
finding.  A similar analysis for the AGASA events above $10^{20}$ eV
yields a KS probability of 0.58 \% comparable to the earlier 0.31 \%
obtained without taking into account any deflection due to magnetic
fields. These results suggest that our conclusions do not depend on
the deflection of cosmic rays due to magnetic fields for the values
discussed in the preceding paragraphs.

\section{Conclusions}

The two dimensional KS test applied to the angular positions of the 60
AGASA events with energies above $4\times 10^{19}$ eV and the angular
positions of the far infrared galaxies in the IRAS PCSz catalog show
that the sources of these particles are consistent with impulsive
events associated with core collapse supernovae.  The KS probability
for this association is $\sim 50$\%, too large to reject the
hypothesis.  Possible physical models of accelerators associated with
such supernovae are relativistic shock acceleration associated with
long burst GRBs and electromagnetic processes in the relativistic
winds from newly born, rapidly rotating magnetars.

However, application of the same test to the 11 AGASA events with
energy above $10^{20}$ eV yields a KS probability of $< 0.5$\%,
showing that the existing set of super GZK events are not consistent
with the GRB model, the magnetar model, or any other model based on
impulsive events associated with core collapse supernovae.  At the
$\sim 2\sigma$ level, this inconsistency is unlikely to have occurred
by chance. However, 2$\sigma$ results in high energy particle and
photon astrophysics are notorious for their lack of
robustness. Therefore, this highest energy result, which uses a very
small number of cosmic ray events, is perhaps best regarded as a
prediction that will be tested by the AUGER experiment. That experiment
will yield enough events above $10^{20}$ eV to show whether the
consistency between these compact object models and the UHE data using
the whole set of results from $4\times 10^{19}$ eV and above, or the
inconsistency between models and data using only the super $10^{20}$
eV data, is the correct conclusion.

\begin{acknowledgments}

C.-P. M. is partially supported by an Alfred P. Sloan fellowship, a
Cottrell Scholars Award from the Research Corporation, and NASA grant
NAG5-12173. J.A. is partially supported by NASA grants NAG 5-12031 and
HST-AR-09548.01-A, and by the Miller Institute for Basic Research in
Science.  We thank California's taxpayers for their indulgence.

\end{acknowledgments}

\clearpage

\begin{figure*}
\begin{tabular}{c}
\epsfig{file=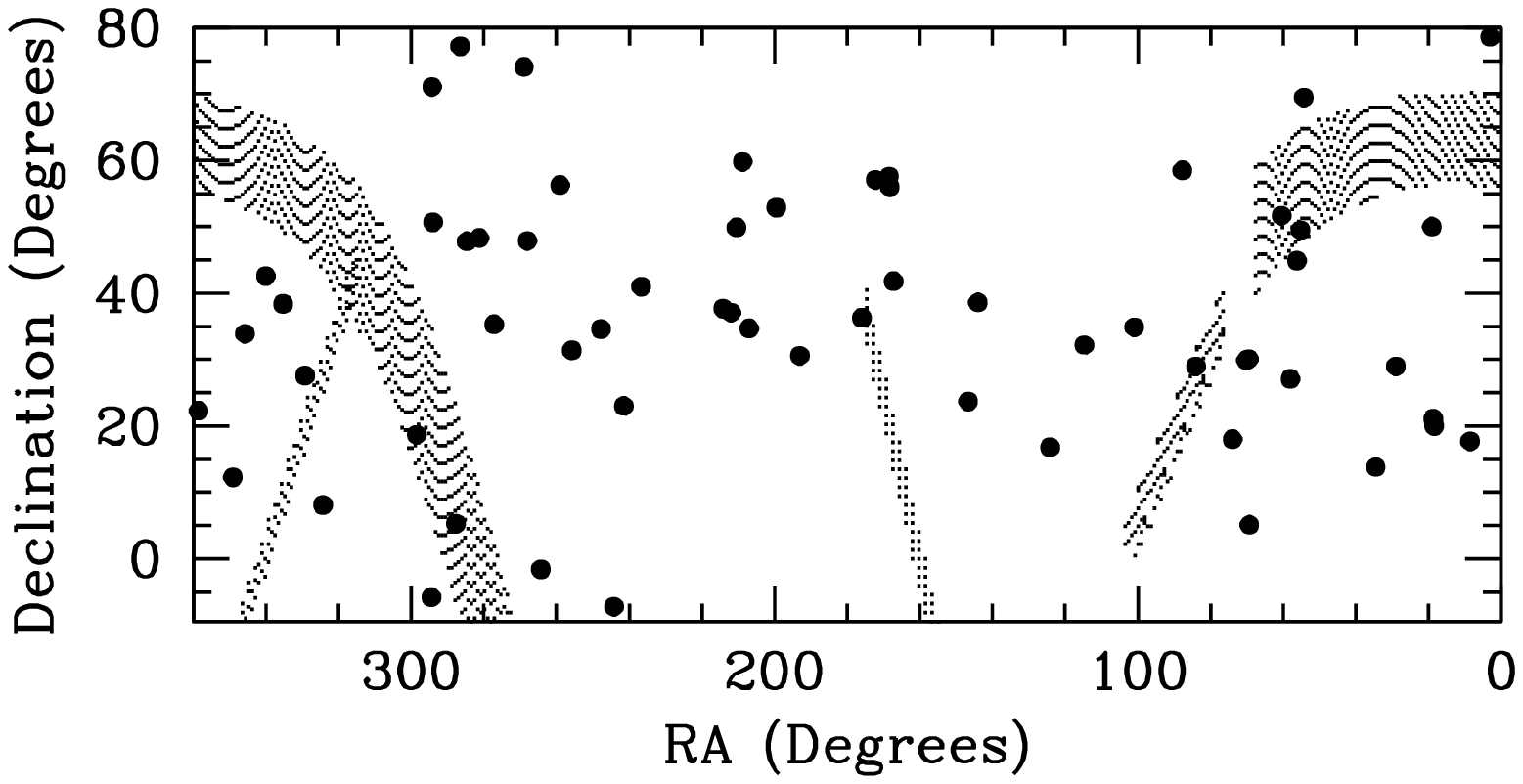}
\end{tabular}
\caption{ Sky distribution of the 60 AGASA data points above $4 \times
10^{19}$ eV (filled circles) and the approximate areas not mapped by IRAS
(black bands).  Six AGASA events lie in the unmapped IRAS region,
which we exclude from our study.  }
\end{figure*}

\begin{figure*}
\begin{tabular}{c}
\epsfig{file=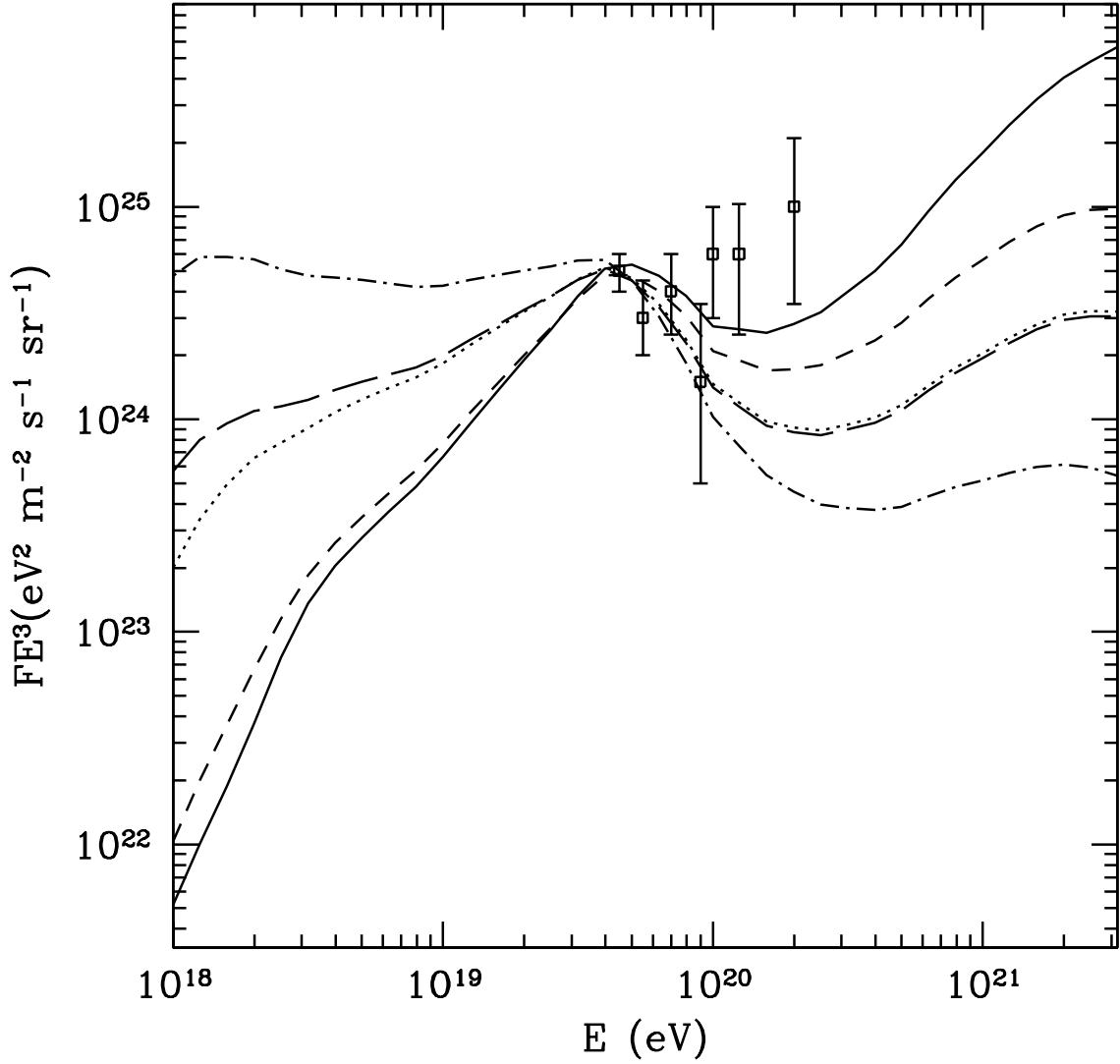,width=7.5in}
\end{tabular}
\caption{ Predictions for the cosmic ray flux as a function of energy
assuming an isotropic distribution of sources with uniform comoving
density for the five different injection spectra analyzed in this
work: magnetar models with no GR loss (solid), moderate GR loss (short
dashed), and strong GR loss (dotted), and GRB models with $E^{-2}$
(long dashed), and $E^{-2.5}$ (dot dashed).  The data points (squares)
are taken from Ref. \cite{TA03} for an earlier set of 30 AGASA UHECR
events.  The GZK suppression is more important for steeper injection
spectra.}
\end{figure*}

\begin{figure*}
\begin{tabular}{c}
\epsfig{file=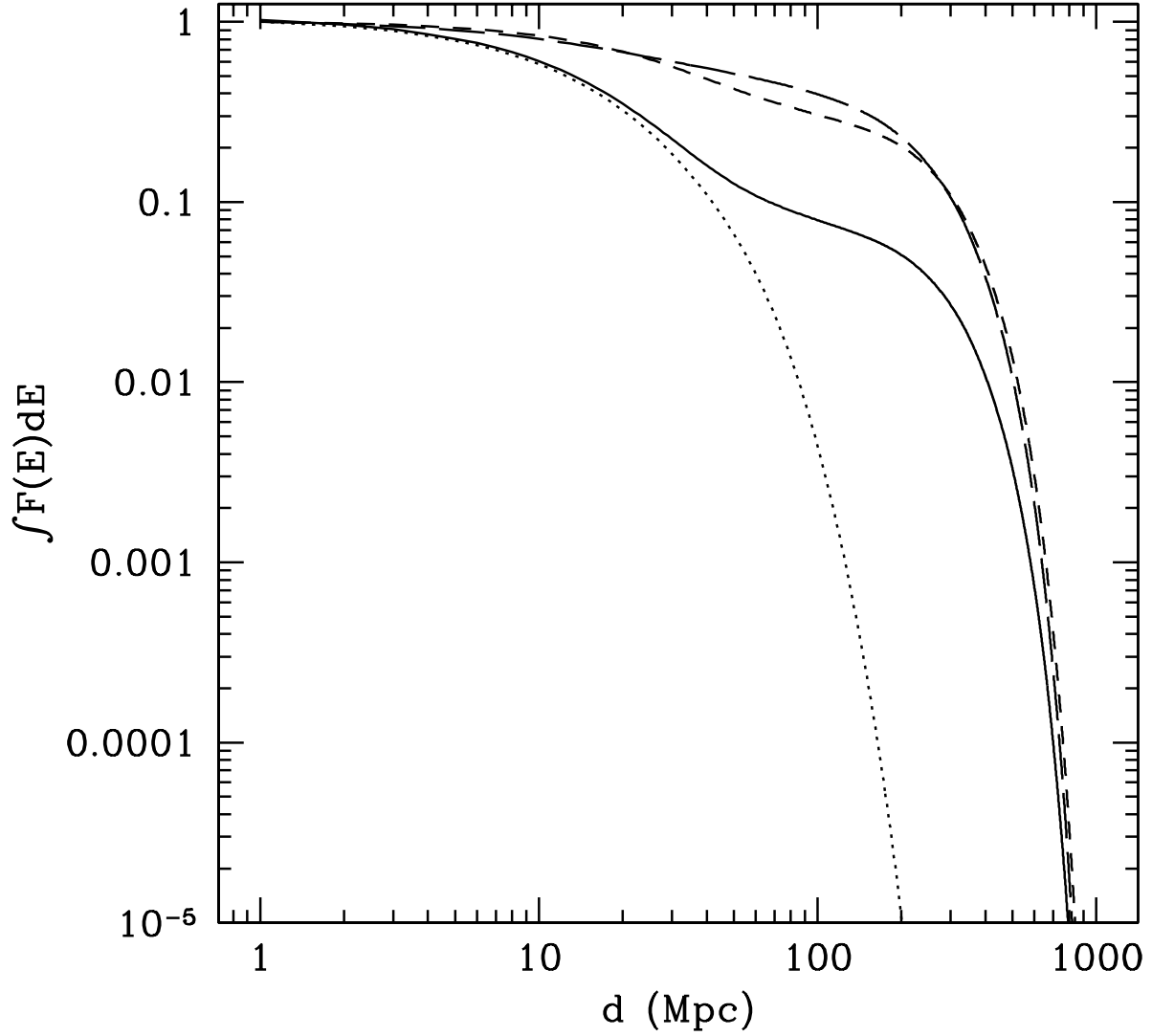,width=7.5in}
\end{tabular}
\caption{ The integrated flux as a function of the source distance for
the no GR loss injection spectrum for $E>4 \times 10^{19}$ eV (solid)
and $E> 10^{20}$ eV (dotted). In the latter case the integrated flux
falls off very sharply for $d> 50$ Mpc. Also shown are the integrated
flux above $E>4 \times 10^{19}$ eV for the moderate GR (short dashed)
and $E^{-2.5}$ (long dashed) injection spectrum. The upper cutoff in
the injection spectrum was taken to be $3 \times 10^{22}$ eV for all
cases in accordance with \cite{A03}.}
\end{figure*}

\begin{figure*}
\begin{tabular}{c}
\epsfig{file=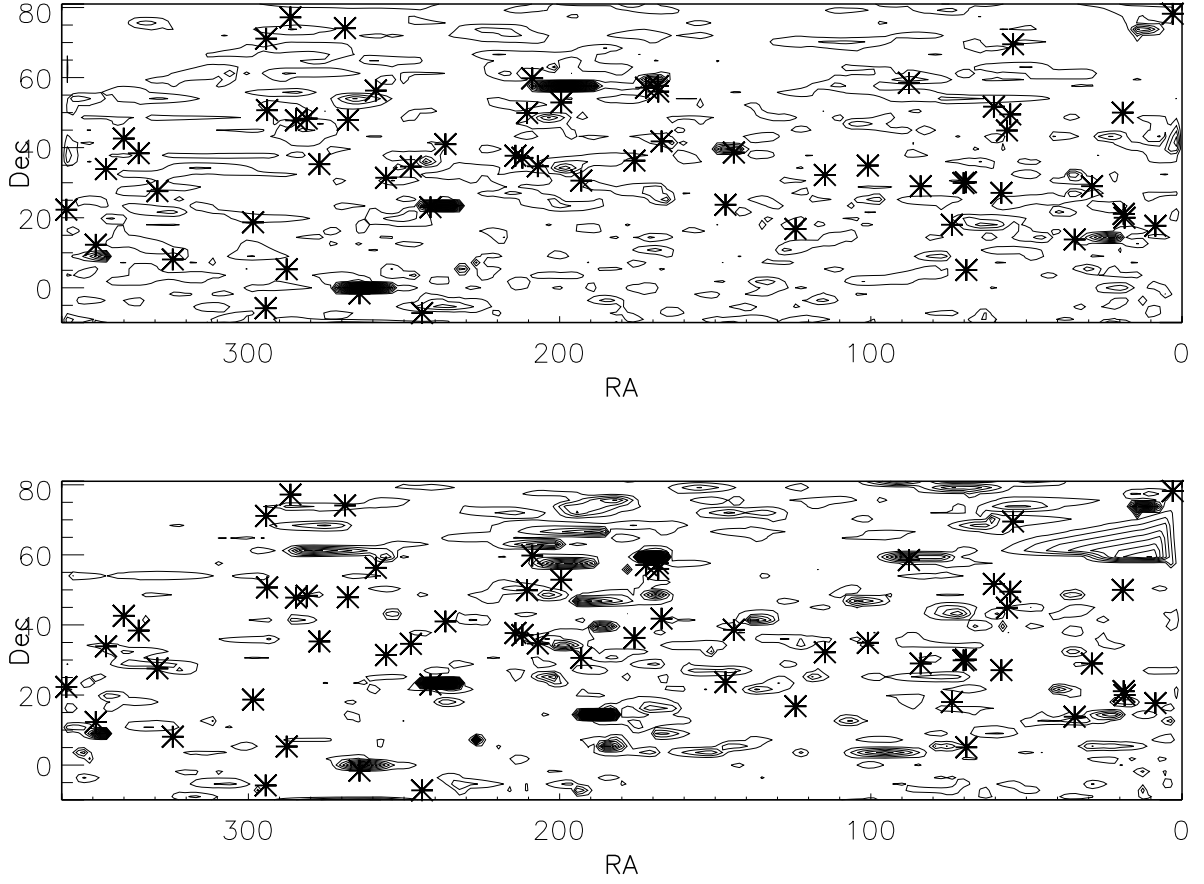}
\end{tabular}
\caption{ Contour plot of the expected integrated cosmic ray flux from
IRAS galaxies smoothed to the AGASA angular resolution of
1.8$^{\circ}$.  The 60 AGASA UHECR events above $4 \times 10^{19}$ eV
are shown as stars.  Two models are shown for comparison: the magnetar
model with no GR loss (top) and the GRB model with $E^{-2}$ injection
spectrum.  No corrections for the incomplete IRAS sky coverage and
non-uniform AGASA detector efficiency in declination (see Sec~III)
have been applied.}
\end{figure*}

\end{document}